\documentclass[pra,dvips,floatfix,superscriptaddress,intlimits,showpacs,twocolumn]{revtex4}

\usepackage{amsfonts,amsmath,amssymb}
\usepackage{dcolumn}
\usepackage{graphicx}

\begin{document}

\title{Low-energy inelastic collisions of OH radicals with He atoms and D$_2$ molecules }%

\author{Moritz Kirste}
\affiliation{Fritz-Haber-Institut der Max-Planck-Gesellschaft,
Faradayweg 4-6, 14195 Berlin, Germany}
\author{Ludwig Scharfenberg}
\affiliation{Fritz-Haber-Institut der Max-Planck-Gesellschaft,
Faradayweg 4-6, 14195 Berlin, Germany}
\author{Jacek K\l os}
\affiliation{Department of Chemistry and Biochemistry, University of
Maryland, College Park, MD 20742-2021, USA}
\author{Fran\c{c}ois Lique}
\affiliation{LOMC, Universit\'e du Havre, 25 Rue Philippe Lebon, BP
540-76 058, Le Havre Cedex, France}
\author{Millard H. Alexander}
\affiliation{Department of Chemistry and Biochemistry and Institute
for Physical Science and Technology, University of Maryland, College
Park, MD 20742-2021, USA}
\author{Gerard Meijer}
\affiliation{Fritz-Haber-Institut der Max-Planck-Gesellschaft,
Faradayweg 4-6, 14195 Berlin, Germany}
\author{Sebastiaan Y.T. van de Meerakker}
\affiliation{Fritz-Haber-Institut der Max-Planck-Gesellschaft,
Faradayweg 4-6, 14195 Berlin, Germany}

\date{\today}%
\begin{abstract}\noindent%
We present an experimental study on the rotational inelastic
scattering of OH ($X^2\Pi_{3/2}, J=3/2, f$) radicals with He and
D$_2$ at collision energies between 100~and~500 cm$^{-1}$ in a
crossed beam experiment. The OH radicals are state selected and
velocity tuned using a Stark decelerator. Relative parity-resolved
state-to-state inelastic scattering cross sections are accurately
determined. These experiments complement recent low-energy collision
studies between trapped OH radicals and beams of He and D$_2$ that
are sensitive to the total (elastic and inelastic) cross sections
(Sawyer \emph{et al.}, \emph{Phys. Rev. Lett.} \textbf{2008},
\emph{101}, 203203), but for which the measured cross sections could
not be reproduced by theoretical calculations (Pavlovic \emph{et
al.}, \emph{J. Phys. Chem. A} \textbf{2009}, \emph{113}, 14670). For
the OH-He system, our experiments validate the inelastic cross
sections determined from rigorous quantum calculations.
\end{abstract}%
\pacs{34.50.-s,34.50.Ez,37.10.Mn}
\maketitle%

\section{Introduction}\label{sec:introduction}
The study of collisions between neutral atoms and molecules at low
collision energies is a fast developing field in molecular physics
\cite{Bell:MolPhys107:99,Smith:AC45:2842,NewJPhys:SpecialIssue,Smith:LowTemperatures,Krems:ColdMolecules}.
This growing interest originates from the exotic scattering
properties of molecules at low temperatures. At temperatures below
$\sim$ 10~K only a few partial waves contribute to the scattering,
leading to dramatic changes in the dynamics. Scattering resonances
can occur when the collision energy is degenerate with a bound state
of the collision complex \cite{Chandler:JCP132:110901}. Low
collision energies also allow for external control over the
collision dynamics by electromagnetic fields. At collision energies
below a few Kelvin, the perturbations due to the Zeeman and Stark
effect become comparable to the translational energy, opening the
possibility for controlled chemistry \cite{Krems:PCCP10:4079}.

In recent years, a variety of novel experimental methods have been
developed that enable scattering experiments at lower collision
energies and/or with a higher precision than hitherto possible. The
buffer-gas cooling, the Stark deceleration, and the velocity
selection techniques have already successfully been applied to
molecular scattering experiments
\cite{Maussang:PRL94:123002,Campbell:PRL102:013003,Gilijamse:Science313:1617,Willitsch:PRL100:043203}.
In the ultra-cold regime, spectacular advances have been made in the
study of interactions between alkali-dimers near quantum degeneracy
\cite{Ospelkaus:Science327:853,Ni:Nature464:1324}. Together with the
collection of other techniques that are currently being developed,
these methods have the potential to start a new era in molecular
scattering experiments.

The Stark deceleration technique has excellent potential for precise
molecular scattering studies as a function of the collision energy.
Compared to conventional molecular beam sources, a Stark decelerator
produces beams of molecules with a narrow velocity spread, perfect
quantum state purity, and with a computer controlled velocity
\cite{Meerakker:NatPhys4:595}. So far, two experimental approaches
have been followed to use these monochromatic beams in molecular
scattering studies.

In 2006, the first scattering experiment using a Stark decelerated
molecular beam was performed. Stark-decelerated OH radicals were
scattered with a supersonic beam of Xe atoms under 90~degree angle
of incidence \cite{Gilijamse:Science313:1617}. This crossed
molecular beam configuration allowed the accurate measurement of the
relative inelastic scattering cross sections as a function of the
collision energy in the collision energy range of 50~to~400
cm$^{-1}$. Recently, this experimental approach was improved
significantly using a superior Stark decelerator
\cite{Scharfenberg:PRA79:023410}. With this decelerator, scattering
experiments can be performed with a better sensitivity, as has been
demonstrated for the benchmark OH($^2\Pi$)-Ar system
\cite{Scharfenberg:PCCP}. In both experiments, excellent agreement
was obtained with cross sections determined by quantum
close-coupling calculations based on high-quality \emph{ab initio}
OH-Xe and OH-Ar potential energy surfaces (PES's).

In another experiment, the approach to confine the Stark-decelerated
OH radicals in a permanent magnetic trap prior to the collision was
followed \cite{Sawyer:PRL101:203203}. Collisions with the OH
radicals were studied by sending supersonic beams of He atoms or
D$_2$ molecules through the trap. Information on the total collision
cross sections could be inferred from the beam-induced trap loss
that occurs through elastic as well as inelastic collisions. The
collision energy was varied from 60~to 230~cm$^{-1}$ for collisions
with He and from 145 to 510 cm$^{-1}$ for collisions with D$_2$ by
changing the temperature of the pulsed solenoid valve used to
produce the supersonic beams. Absolute collision cross sections were
determined by calibrating the beam flux using a pressure
measurement. Indications for quantum threshold scattering at a
collision energy of 84 cm$^{-1}$, equal to the energy splitting of
the two lowest lying rotational levels of the OH radical, were found
for collisions between OH and He. For OH-D$_2$ collisions, a
pronounced peak in the total cross section was observed at collision
energies around 305~cm$^{-1}$, an energy that is equal to the energy
splitting between the $J=1$ and the $J=3$ rotational levels of
para-D$_2$. The enhancement of the cross section at this energy was
tentatively attributed to resonant energy transfer between the OH
radicals and D$_2$ molecules.

Both experimental approaches have been successful in demonstrating
the feasibility of using Stark decelerated molecules in scattering
experiments. The experimental results of the latter experiment,
however, could not be reproduced by theoretical calculations. The
decrease in the total cross section that was observed below
84~cm$^{-1}$ for the scattering between OH and He was not reproduced
by rigorous quantum calculations for low temperature collisions of
OH-He. The calculations show that the total cross section
\emph{increases} significantly at collision energies below 100
cm$^{-1}$ \cite{Pavlovic:JPCA113:14670}. The experimentally observed
threshold behavior can be explained if the trap loss originates
mainly from inelastic scattering, although this appears unlikely for
the kinetic conditions of the experiment \cite{Tscherbul:unpubl}.
The measured cross section of $\sim$ 100 \AA$^2$ at a collision
energy of 150 cm$^{-1}$ is an order of magnitude larger than the
theoretical total inelastic cross section, indicating that the
majority of the collisions that lead to trap loss are indeed
elastic. The source of the discrepancy between theory and
experiment, and whether or not the presence of the trapping field in
the collision experiment can explain the observed cross sections, is
at present unclear. Unfortunately, no rigorous quantum calculations
are available for the OH-D$_2$ system to compare calculated cross
sections with experimental ones, or to investigate the physical
origin of the intriguing peak that was observed at a collision
energy of 305 cm$^{-1}$.

Here we complement the low-energy collision studies between OH-He
and OH-D$_2$ of Sawyer \emph{et al.} \cite{Sawyer:PRL101:203203} by
investigating rotational energy transfer in collisions of
Stark-decelerated OH ($X\,^2\Pi_{3/2}, J=3/2, f$) radicals with He
atoms and D$_2$ molecules in a crossed beam experiment under field
free conditions. The OH-He and OH-D$_2$ center-of-mass collision
energies are tuned from 120 cm$^{-1}$ to 400 cm$^{-1}$ and from 150
cm$^{-1}$ to 500 cm$^{-1}$, respectively. Parity-resolved
state-to-state relative inelastic scattering cross sections are
accurately measured. For the OH-He system, good agreement is
obtained with the inelastic cross sections determined by
close-coupled calculations based on the OH-He PES's used in Ref.
\cite{Pavlovic:JPCA113:14670}, validating the theoretical
predictions for the low-energy inelastic scattering between OH
radicals and He atoms. For the OH-D$_2$ system, no strong variation
in the state-to-state relative inelastic scattering cross sections
is found at center-of-mass energies around 300 cm$^{-1}$.

The interaction of OH radicals with, in particular, He atoms has
been the subject of extensive experimental and theoretical
investigations, and is of direct astrophysical relevance. The
rotational energy transfer of OH by collisions with H$_2$ molecules
in interstellar clouds is believed to play an important role in the
formation of interstellar OH masers \cite{Elitzur:RMP54:1225}. The
(mass scaled) collision cross sections for the theoretically more
tractable OH-He system are often used to model the collision
dynamics in these environments. In a crossed beam experiment,
Schreel \emph{et al.} prepared the OH radicals in the upper
$\Lambda$-doublet component of the $X\,^2\Pi_{3/2}, J=3/2$ level by
hexapole state selection \cite{Schreel:JCP99:8713}. Accurate
state-to-state inelastic scattering cross sections were obtained at
a collision energy of $\sim$ 400 cm$^{-1}$. The effect of
vibrational excitation of the OH radical on the rotational energy
transfer has been investigated by Wysong \emph{et al.}
\cite{Wysong:JCP94:7547}. The bound states of the weakly bound OH-He
complex were spectroscopically investigated by Han and Heaven
\cite{Han:JCP123:064307}. The depolarization of rotationally excited
OH radicals with He under thermal conditions has been studied using
two-color laser spectroscopy by Paterson \emph{et al.}
\cite{Paterson:JCP129:074304}. All these experiments have been in
good agreement with theoretical calculations based on accurate
\emph{ab initio} potential energy surfaces
\cite{DegliEsposti:JCP103:2067,Lee:JCP113:5736,Dagdigian:JCP130:164315}.
The interaction between OH radicals and D$_2$ molecules has been
studied less extensively. In crossed beam experiments, inelastic as
well as reactive scattering processes have been studied at high
collision energies
\cite{Andresen:JCP95:5763,Alagia:CP207:389,Strazisar:Science290:958}.
The system has recently been treated theoretically using the
quasi-classical trajectory method \cite{Sierra:PCCP11:11520}.
Rotational inelastic cross sections for the related OH-H$_2$ system
have been calculated by Van Dishoeck and coworkers
\cite{Offer:JCP100:362}.

\section{Experiment}
\begin{figure}[!htb]
    \centering
    \resizebox{1.0\linewidth}{!}
    {\includegraphics[0,0][450,380]{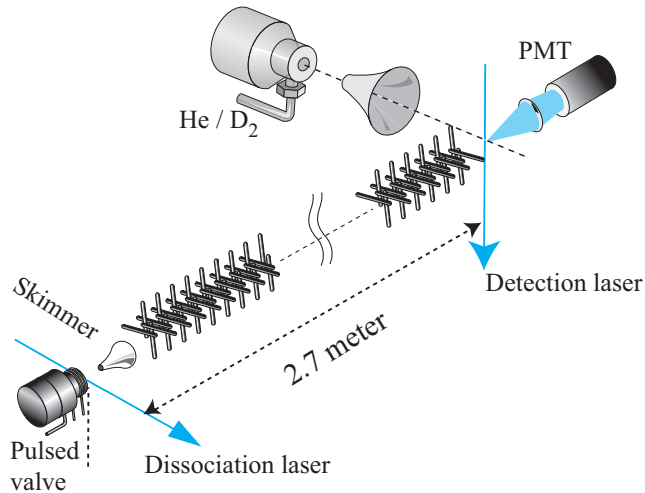}}
    \caption{Scheme of the experimental setup. A pulsed beam of OH radicals
    is produced via photodissociation of HNO$_3$ seeded in an inert carrier
    gas. The OH radicals pass through a 2.6-m-long Stark decelerator,
    and are scattered with a pulsed beam of He atoms or D$_2$ molecules. The OH radicals
    are state-selectively detected via laser-induced fluorescence, that is recorded with a photomultiplier tube (PMT).}
    \label{fig:setup}
\end{figure}

The experiments are performed in a crossed molecular beam machine
that is schematically shown in Figure \ref{fig:setup}. This machine
has been used recently to study the rotational energy transfer in
collisions between state-selected OH ($X\,^2\Pi_{3/2}, J=3/2, f$)
radicals and Ar atoms as a function of the collision energy
\cite{Scharfenberg:PCCP}. A detailed description of the production,
Stark deceleration and detection of the OH radicals, as well as of
the procedure that is followed to tune the collision energy is given
in ref. \cite{Scharfenberg:PCCP}; we here limit ourselves to a brief
summary.

A pulsed supersonic beam of OH radicals in the $X\,^2\Pi_{3/2},
J=3/2$ state is produced by photolysis of nitric acid seeded in an
inert carrier gas. The OH radicals that reside in the upper
$\Lambda$-doublet component of $f$ parity are decelerated, guided,
or accelerated with the use of a 2.6~meter long Stark decelerator
that is used in the so-called $s=3$ operation mode
\cite{Meerakker:PRA71:053409,Scharfenberg:PRA79:023410}. The OH
radicals are scattered with a neat beam of He atoms or D$_2$
molecules at a distance of 16.5~mm from the exit of the decelerator.
The beams scatter under 90$^{\circ}$ angle of incidence and
collisions take place in a field free region. The density of the He
and D$_2$ molecular beams are kept sufficiently low to ensure single
collision conditions.

\begin{figure}[!htb]
    \centering
    \resizebox{1.0\linewidth}{!}
    {\includegraphics[0,0][600,500]{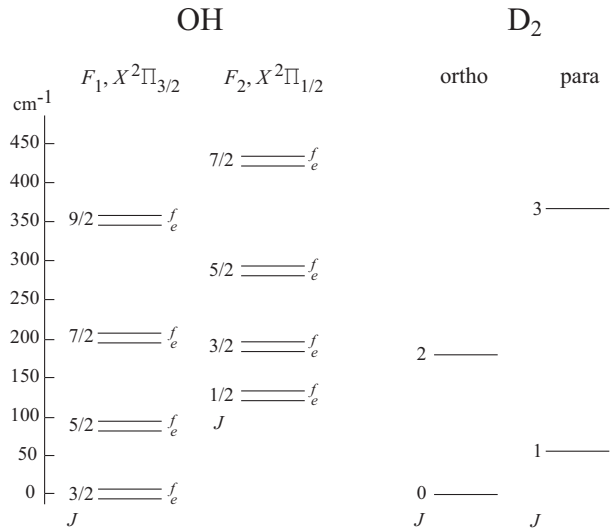}}
    \caption{Rotational energy level diagram of the OH radical (left)
    and the D$_2$ molecule (right). The rotational states are labeled
    with the rotational quantum number $J$. The spectroscopic symmetry
    labels $e$ and $f$ are used to denote the two $\Lambda$-doublet components
    that exist for every rotational state of OH. The splitting between both
    components is largely exaggerated for reasons of clarity.}
    \label{fig:energy-level-scheme}
\end{figure}

The beam of D$_2$ molecules is produced using a gas of normal D$_2$,
and both ortho-D$_2$ and para-D$_2$ molecules contribute therefore
to the measured state-to-state inelastic cross sections. According
to the statistical weights, 67\% and 33\% of the D$_2$ molecules are
expected to reside in a rotational state belonging to ortho and
para-D$_2$, respectively. The rotational energy level diagram of
D$_2$ is schematically shown in Figure
\ref{fig:energy-level-scheme}. The energy level spacing between the
lowest rotational states is large, and the D$_2$ molecules are
expected to predominantly reside in the $J=0$ and the $J=1$ levels
under our experimental conditions.

The collision energy is varied by tuning the velocity of the OH
radicals, and by choosing different temperatures for the solenoid
valve used to produce the He and D$_2$ beams. The mean forward
velocity of the He/D$_2$ beam is measured by two microphone based
beam detectors placed 300~mm apart. For a given temperature of the
valve, similar beam speeds are measured for He and D$_2$ as is to be
expected for beam particles of identical mass. The lowest collision
energies are obtained when the solenoid valve is cooled to near
liquid nitrogen temperatures, resulting in a minimum mean beam
velocity of 996~m/s for He and 1042~m/s for D$_2$. These velocities
are expected for particles with a mass of 4~atomic units and near
liquid nitrogen nozzle temperatures. The slightly higher speeds that
are measured for D$_2$ beams are attributed to the extra degrees of
freedom of the D$_2$ rotor compared to He.

The lowest collision energy amounts to 120~cm$^{-1}$ and
150~cm$^{-1}$ for OH-He and OH-D$_2$, respectively. In the
experiment by Sawyer \emph{et al.}, a minimum collision energy of
$\sim$ 60~cm$^{-1}$ for OH-He and $\sim$ 145~cm$^{-1}$ for OH-D$_2$
was obtained \cite{Sawyer:PRL101:203203}. It is noted that the
significantly lower collision energy for OH-He that was reached in
that experiment is not due to the use of trapped OH radicals. A
center of mass collision energy of $\sim$ 60~cm$^{-1}$ for
collisions between stationary OH radicals and helium atoms requires
a He atom velocity of $\sim$ 670 m/s. This is much lower than the
expected velocity for a supersonic beam of helium atoms at the
temperature used, and the atomic beam that was employed by Sawyer
\emph{et al.} is believed to have been effusive-like
\cite{Sawyer:pricom2009}.

Collision energies up to 400~cm$^{-1}$ for OH-He and 500~cm$^{-1}$
for OH-D$_2$ are reached by tuning the velocity of the OH radicals
between 168 and 741~m/s, and by using temperatures of 293~K, 253~K,
213~K, 173~K, 133~K, and 93~K for the valve producing the He/D$_2$
beam. The width (full width at half maximum) of the collision energy
distribution depends on the collision energy, and ranges from $\sim$
20 cm$^{-1}$ at the lowest collision energies to $\sim$ 60 cm$^{-1}$
at the highest collision energies.

Collisional excitation of the OH radicals up to the $X\,^2\Pi_{3/2},
J=9/2$ and the $X\,^2\Pi_{1/2}, J=7/2$ state is measured. These
rotational states are schematically shown in the rotational energy
level diagram in Figure \ref{fig:energy-level-scheme}, and are
referred to hereafter as $F_i(Je/f)$, where $i=1$ and $i=2$ are used
to indicate the $X\,^2\Pi_{3/2}$ and the $X\,^2\Pi_{1/2}$ spin-orbit
manifolds, respectively. The inelastically scattered OH radicals are
state-selectively detected via saturated laser induced fluorescence
using different rotational transitions of the $A\,^2\Sigma^{+}, v=1
\leftarrow X\,^2\Pi, v=0$ band. The off-resonant fluorescence is
collected at right angles and imaged into a photomultiplier tube
(PMT). The diameter of the laser beam is approximately 8~mm,
providing a detection volume that is larger than the intersection
volume of both beams.

The experiment runs at a repetition rate of 10~Hz. The beam that
provides the collision partner is operated every second shot of the
experiment, and the collision signal results as the signal intensity
difference of alternating shots of the experiment. The collision
energy is varied in a quasi-continuous cycle, as has been described
in detail in ref. \cite{Scharfenberg:PCCP}. For the strongest
scattering channels, the fluorescence signals are recorded using an
analog mode of detection; the weak signals are analyzed using photon
counting. Both modes of signal acquisition are calibrated with
respect to each other.

To relate the measured signal intensities to collision induced
populations, the different excitation rates for the different
branches of the optical transitions used to probe the different
rotational levels are taken into account in the data analysis
\cite{Scharfenberg:PCCP}. The measured relative populations in the
various rotational states directly reflect inelastic cross sections.
No density-to-flux transformation is required for crossed beam
scattering experiments using a light particle as collision partner
\cite{Sonnenfroh:CPL176:183}. The validity of this assumption is
verified by a measurement of the variation of the relative collision
signals as a function of time in the overlapping beams. No variation
was recorded in the experiment, in agreement with model calculations
of the detection probabilities of the scattered molecules.

We now describe the theoretical methods used to calculate the cross
sections for the inelastic scattering between OH radicals and He
atoms; a discussion of the experimental relative cross sections is
given in section \ref{sec:discussion}.

\section{Theory}\label{sec:theory}
Fully quantum, close-coupling scattering calculations of inelastic
collisions of OH radicals with He atoms have been performed recently
by K{\l}os \emph{et al.} in Ref. \cite{Klos:CPL445:12} based on the
RCCSD(T) potential energy surfaces of Lee \emph{et al.}
\cite{Lee:JCP113:5736}. K{\l}os \emph{et al.} in Ref.
\cite{Klos:CPL445:12} present cross sections for rotational
transitions out of the $F_1(3/2e)$ state; for the present experiment
transitions out of the $F_1(3/2f)$ states are of relevance. Below we
briefly describe the scattering methodology relevant for the OH-He
system and some of the details of the calculations presented in Ref.
\cite{Klos:CPL445:12}.

The interaction between the open shell OH($X\,^2\Pi$) radical and a
spherical He atom is described by two PES's $V_{A'}$ and $V_{A''}$,
having $A'$ and $A''$ reflection symmetry in the plane containing
the OH radical and the He atom \cite{Alexander:ChemPhys92:337}. The
PES of $A'$ and $A''$ symmetry describes the OH-He interaction where
the OH radical has its singly occupied $\pi$ orbital in and
perpendicular to the triatomic plane, respectively. In scattering
calculations it is more convenient to construct the average
potential $V_{\textrm{sum}}=1/2(V_{A''}+V_{A'})$ and the
half-difference potential $V_{\textrm{dif}}=1/2(V_{A''}-V_{A'})$ of
these PES's \cite{Alexander:JCP76:5974,Alexander:ChemPhys92:337}.
The HIBRIDON program suite was used to carry out fully-quantum,
close-coupling calculations of integral state-to-state scattering
cross sections \cite{Hibridon:hib43}. The channel basis was chosen
to ensure convergence of the integral cross sections for all $J,F_i
\rightarrow J',F'_i$ transitions with $J,J'\le$ 11.5. The calculated
inelastic cross sections were converged to within 0.01 \AA$^2$.

\section{Results and Discussion} \label{sec:discussion}
\begin{figure}[!htb]
    \centering
    \resizebox{1.0\linewidth}{!}
    {\includegraphics[0,0][340,400]{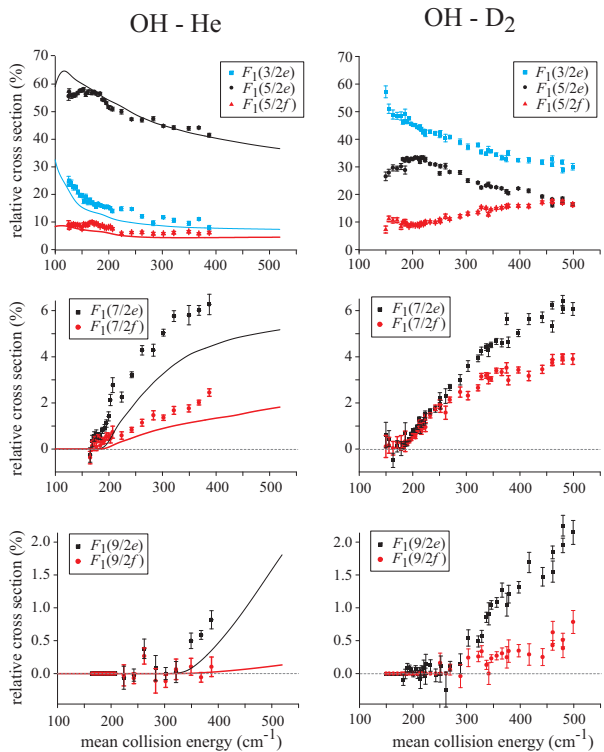}}
    \caption{\emph{(Color online)} Relative state-to-state inelastic
    scattering cross sections for spin-orbit conserving ($F_1 \rightarrow F_1$) collisions of OH ($X\,^2\Pi_{3/2}, J=3/2, f$)
    radicals with helium atoms (left) and D$_2$ molecules (right).
    The theoretically calculated cross sections for the scattering of OH with He from Ref. \cite{Klos:CPL445:12} are included as solid curves in the left panels.}
    \label{fig:results_F1}
\end{figure}

\begin{figure}[!htb]
    \centering
    \resizebox{1.0\linewidth}{!}
    {\includegraphics[0,0][340,400]{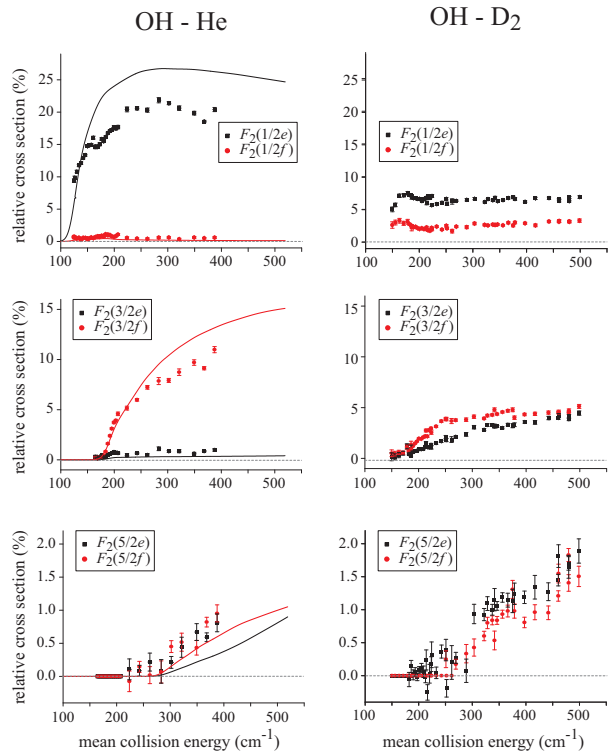}}
    \caption{\emph{(Color online)} Relative state-to-state inelastic
    scattering cross sections for spin-orbit changing ($F_1 \rightarrow F_2$) collisions of OH ($X\,^2\Pi_{3/2}, J=3/2, f$)
    radicals with helium atoms (left) and D$_2$ molecules (right).
    The theoretically calculated cross sections for the scattering of OH with He from Ref. \cite{Klos:CPL445:12} are included as solid curves in the left panels.}
    \label{fig:results_F2}
\end{figure}

In Figures \ref{fig:results_F1} and \ref{fig:results_F2}, the
measured relative state-to-state inelastic scattering cross sections
to levels in the $F_1$ manifold (spin-orbit conserving collisions)
and the $F_2$ manifold (spin-orbit changing collisions),
respectively, are shown. The cross sections that are obtained for
the collision partners He and D$_2$ are displayed in the left and
right hand side of these Figures. The scattering channels that
correspond to excitation of the OH radicals to the two different
$\Lambda$-doublet components of a given rotational state are grouped
together. In the upper panels, the scattering channel that populates
the $F_1(3/2e)$ state is shown together with the channels that
populate both $\Lambda$-doublet components of the $F_1(5/2)$
rotational state. To facilitate a direct comparison between the
scattering cross sections for OH-He and OH-D$_2$, identical axes are
used in the panels that correspond to the same scattering channels.
The theoretically calculated cross sections for the scattering of OH
with He, convoluted with the experimental energy resolution, are
included as solid curves in the left panels.

\subsection{OH-He}
In the collision energy range that is probed, the rotational
inelastic scattering of OH $(F_1(3/2f))$ radicals with He atoms is
dominated by excitation to the $F_1(5/2e)$ state. The $F_1(3/2e)$
channel, corresponding to collisions that induce the $J=3/2, f
\rightarrow J=3/2, e$ $\Lambda$-doublet transition in the OH
radical, appears rather weak. This is in contrast with the
scattering of OH radicals with Ar and Xe atoms for which the
$F_1(3/2e)$ channel is the dominant inelastic scattering channel
\cite{Gilijamse:Science313:1617,Scharfenberg:PCCP}. The scattering
channel populating the $F_2(1/2e)$ state appears exceptionally
large, also at variance with the corresponding cross sections for
the collision partners Ar and Xe.

For spin-orbit manifold conserving collisions, there is a strong
propensity for final states of $e$ parity. For spin-orbit manifold
changing collisions populating the $J=1/2$ and $J=3/2$ states, very
strong propensities are observed, showing a near symmetry selection
rule. Collisions that populate the $F_2(1/2e)$ and $F_2(3/2f)$
states are approximately two orders of magnitude more effective than
collisions populating the $F_2(1/2f)$ and $F_2(3/2e)$ states,
respectively.

At high collision energies, the relative state-to-state cross
sections and the propensities are consistent with the observations
by Schreel \emph{et al.} \cite{Schreel:JCP99:8713}. In that
experiment, however, the strong propensities were partially
concealed due to a sizable initial population in the $F_1(5/2f)$
state \cite{Schreel:JCP99:8713}. The almost perfect quantum state
purity of the packets of OH radicals that are used in the present
experiment enables to unambiguously measure the cross sections of
transitions to final states that are only weakly coupled to the
$F_1(3/2f)$ initial state.

Throughout the range of collision energies, a good agreement is
found with the computed cross sections. The relative scattering
cross sections and the propensities for transitions to one of the
$\Lambda$-doublet components of the final rotational state, are
reproduced well. The largest difference between theory and
experiment is found for the $F_2(1/2e)$ channel.

It is interesting to investigate the physical origin of the general
behavior of the scattering cross sections and in particular the
$\Lambda$-doublet propensities. A general picture of the scattering
of $^2\Pi$ molecules with spherical objects has been developed by
Dagdigian \emph{et al.} \cite{Dagdigian:JCP91:839}, and has been
applied to the inelastic scattering of OH radicals with atomic
collision partners before
\cite{DegliEsposti:JCP103:2067,Beek:JCP113:628}. From the formal
quantum analysis of the scattering, it follows that the coupling
between rotational states can be evaluated from
\begin{equation}\label{eq:matrixelement}
\sum_{l} K^{l} \left[A^{l}_{\Omega' J' \epsilon', \Omega J \epsilon}
V_{l0} + B^{l}_{\Omega' J' \epsilon', \Omega J \epsilon} V_{l2}
\right]
\end{equation}
where $J$ is the rotational quantum number of the OH radical,
$\epsilon$ is the symmetry index of the rotational state, and
$\Omega$ is the projection of $J$ onto the internuclear axis. Primed
quantum numbers indicate the post-collision conditions. The terms
$V_{l0}$ and $V_{l2}$ are the expansion coefficients of the average
and difference potentials in terms of regular and associated
Legendre polynomials, respectively. The sum in equation
(\ref{eq:matrixelement}) is performed over the expansion index $l$.

The factor $K^{l}$ is only nonzero for states that are coupled by
the interaction potential, and needs for our analysis no further
discussion. Essential in the understanding of the inelastic cross
sections are the values for the $V_{l0}$ and the $V_{l2}$
coefficients, and the role of the $A^{l}$ and $B^{l}$ factors. Both
$A^{l}$ and $B^{l}$ are independent of the interaction potential,
and are determined exclusively by the rotational energy level
structure of the molecule. The values of $A^{l}$ and $B^{l}$ for OH
radicals in the $X\,^2\Pi_{3/2}, J=3/2, f$ level are tabulated in
Ref. \cite{Dagdigian:JCP91:839}.

For a pure Hund's case (a) molecule, the values for $B^{l}$ are zero
for spin-orbit manifold conserving collisions, whereas the factors
$A^{l}$ are zero for spin-orbit manifold changing collisions.
Consequently, spin-orbit conserving and spin-orbit changing
transitions are induced exclusively by $V_{\textrm{sum}}$ and
$V_{\textrm{dif}}$, respectively. Within each manifold, $e/f$ parity
changing collisions are governed by the terms for which $\Delta J +
l=$ odd, while $e/f$ parity conserving collisions are described by
the $\Delta J + l=$ even terms. The propensities for preferred
excitation to the $e$ or $f$ component of a final rotational state
originate from the different values for the relevant products
$A^{l}V_{l0}$ or $B^{l}V_{l2}$ that govern these transitions.

For molecules like OH that cannot be described by a pure Hund's case
(a) coupling scheme, both the factors $A^{l}$ and $B^{l}$ are
nonzero, and interference between the average and difference
potentials occurs. The $X^2\Pi$ electronic ground state of the OH
radical originates from a $\pi^3$ electron occupancy, leading to
predominantly positive values for $V_{\textrm{sum}}$ and
$V_{\textrm{dif}}$
\cite{Dagdigian:JCP91:839,DegliEsposti:JCP103:2067}. As a result,
final states for which the dominant $A^{l}$ and $B^{l}$ factors have
equal signs are coupled more strongly to the $F_1(3/2f)$ initial
state than final states for which $A^{l}$ and $B^{l}$ have opposite
signs. The former is the case for the final states of $e$ parity in
the $F_1$ spin-orbit manifold, and for final states of $f$ parity in
the $F_2$ manifold, contributing to the observed propensities. The
anomalous propensity that is observed for the $F_2(1/2e)$ state can
be understood from equation (\ref{eq:matrixelement}). Excitation
into the $F_2(1/2f)$ state is governed exclusively by $V_{10}$ in
combination with a small value for $A$, while the $V_{22}$ term in
combination with a large value for $B$ dominates the excitation into
the $F_2(1/2e)$ state
\cite{DegliEsposti:JCP103:2067,Dagdigian:JCP91:839}.

The measured relative state-to-state cross sections directly yield
qualitative information on the expansion coefficients of the
potential energy surfaces that govern the scattering between OH and
He. Collisions that populate the $F_1(3/2e)$ and $F_1(5/2f)$ states
are governed by the coefficients for which $l=$ odd, while the cross
sections for excitation to the $F_1(5/2e)$ state is governed by the
$l=$ even coefficients. The observed ratio of the state-to-state
cross sections indicates that the leading $l=$ even terms $V_{20}$,
$V_{40}$, and $V_{22}$ contribute significantly to the interaction
potential. The relatively large cross section for spin-orbit
changing collisions populating the $F_2(1/2e)$ state, as well as the
strong propensities that are generally observed, suggests that the
$V_{22}$ coefficient of the difference potential plays a significant
role in the scattering between OH radicals and He atoms. These
effects can be rationalized by the nature of the OH-He interaction
potential. The interaction between OH and He is rather weak and the
anisotropy of the potential energy surface is small. The $V_{A''}$
PES has two shallow and almost equally deep potential wells for the
collinear OH-He and the HO-He geometry, with well depths of
27~cm$^{-1}$ and 22~cm$^{-1}$, respectively \cite{Lee:JCP113:5736}.
Consequently, the $l=$ even coefficients that describe the head-tail
symmetric parts of the potential energy surfaces contribute
significantly to the scattering.

\subsection{OH-D$_2$}
The inelastic scattering of OH radicals with D$_2$ molecules shows
interesting differences compared to OH-He. The largest cross section
is observed for collisions that populate the $F_1(3/2e)$ state. For
the channels that populate the $F_1(5/2)$ states, a propensity for
the $\Lambda$-doublet component of $e$ symmetry is observed for low
collision energies, that vanishes for collision energies of about
500 cm$^{-1}$. The other scattering channels show only modest
propensities. The spin-orbit changing collisions appear rather weak.

Although the formalism that was applied above to the scattering of
OH with He does not strictly apply to non-spherical collision
partners, we can use the formalism to obtain a physical
interpretation of the differences between the scattering of OH with
He and D$_2$. This comparison is particularly interesting, as both
collision partners have equal mass, and mass effects in the dynamics
cancel. The interaction of D$_2$ molecules with OH radicals is
stronger and induction forces are more important than for the
interaction between OH and He. This suggests that the PES is less
head-tail symmetric in comparison to OH-He, and the coefficients of
the potential for which $l=$ odd gain importance compared to the
$l=$ even coefficients. This results in a larger cross section for
$\Lambda$-doublet changing collisions populating the $F_1(3/2e)$
state, smaller spin-orbit changing transitions, and less pronounced
propensities for preferred excitation to one of the two components
of a $\Lambda$-doublet. Similar effects have been observed in
state-to-state inelastic scattering experiments of OH radicals with
polar collision partners such as CO$_2$ and HCl
\cite{Beek:JCP109:1302,Cireasa:JCP122:074319}.

In the relative inelastic scattering cross sections, no effect is
seen from the internal rotational degrees of freedom of the D$_2$
molecule. In particular, no strong variation of the cross sections
at collision energies around 300 cm$^{-1}$ is observed, that could
be indicative of resonant energy transfer between the OH and the
D$_2$ rotors \cite{Sawyer:PRL101:203203}.

\section{Conclusions}
We have presented measurements of the state-to-state rotational
inelastic scattering of Stark-decelerated OH ($X\,^2\Pi_{3/2},
J=3/2, f$) radicals with He atoms and D$_2$ molecules in the
100-500~cm$^{-1}$ collision energy range. The collision energy
dependence of the relative inelastic scattering cross sections,
the threshold behavior of inelastic channels, and the energy
dependence of the state-resolved propensities are accurately
determined. For the scattering of OH with He, good agreement is
found with the inelastic scattering cross sections determined from
quantum close-coupled scattering calculations based on
high-quality \emph{ab initio} OH-He PES's.

The almost perfect quantum state purity of the Stark-decelerated
packets of OH radicals eliminates the contamination of the
scattering signals by initial populations in excited rotational
states. This facilitates a quantitative study of collision induced
transitions to states that are only weakly coupled to the initial
state, and enables the observation of the exceptionally strong
propensities for the inelastic scattering between OH radicals and He
atoms. The genuine relative state-to-state inelastic scattering
cross sections that are measured allow for a more accurate
comparison with computed cross sections than hitherto possible.

Significant differences are found between the inelastic scattering
of OH-He and OH-D$_2$. Although no rigorous quantum calculations
have been performed for OH-D$_2$, these differences can be
understood from the different nature of the OH-D$_2$ interaction
potential. No effect of the rotational degrees of freedom of the D$_
2$ molecule has been observed in the relative inelastic scattering
cross sections.

Our measurements on the low-energy scattering between OH radicals
and He/D$_2$ complement recent scattering experiments in which
pulsed beams of He/D$_2$ are directed through a sample of
magnetically trapped OH radicals. In that experiment, inelastic
collisions, as well as elastic collisions that impart a sufficient
amount of kinetic energy to the OH radicals, lead to a reduced
number of OH radicals in the trap. Recent quantum scattering
calculations, based on the same PES's as used in the present work,
have not been able to reproduce the trap loss observations for
OH-He, however. The state-to-state experiments reported in the
present article are not sensitive to elastic scattering, but
validate the field-free inelastic scattering cross sections for
OH-He determined from quantum scattering calculations. The
theoretical description of low-energy inelastic collisions of these
elementary systems is thus adequate; more experimental and
theoretical work is needed to find the source of the discrepancy
between theory and beam-trap experiments. For the accurate
interpretation of collision experiments using trapped molecules, it
is essential to establish the role of elastic scattering and the
influence of the trapping potential on the measured trap loss.

\section{Acknowledgements}
This work is supported by the ESF EuroQUAM programme, and is part of
the CoPoMol (Collisions of Cold Polar Molecules) project. The expert
technical assistance of Henrik Haak, Georg Hammer, and the FHI
mechanical and electronics workshops are gratefully acknowledged. We
thank Timur Tscherbul for stimulating discussions. The theoretical
calculations were supported  by the U. S. National Science
Foundation, grant number CHE-0848110.


\end{document}